\documentclass[aps,showpacs]{revtex4}
\usepackage{amsmath}
\usepackage{graphicx}
\begin{document}
\bibliographystyle{apsrev}


\title{Inadequacy of Scaling Arguments for Neutrino Cross Sections  
}
\author{A. C. Hayes}
\affiliation{Theoretical Division, Los Alamos National Laboratory, 
Los Alamos, New Mexico 87545}
\date{\today}

\begin{abstract}
The problem with the use of scaling arguments 
for simultaneous studies
of different  weak interaction processes is discussed. 
When different 
neutrino scattering cross sections involving quite
different momentum transfers are being compared
it difficult to define a meaningful single
scaling factor to renormalize calculated cross sections. 
It has been suggested that the use of such scaling can be used to estimate
high energy neutrino cross sections from low energy neutrino cross sections.
This argument has lead to questions on the consistency of the
magnitude of the LSND muon neutrino cross sections on $^{12}$C
relative to other lower energy
weak processes.
The issue is revisited here and from inspection of the structure of the form factors involved
it is seen that
the problem arises from a  poor description of the 
transition form factors at high momentum transfer.
When wave functions that reproduce the transverse magnetic
inelastic (e,e') scattering  form factor for the 15.11 MeV state in $^{12}$C are used 
there is no longer a need for scaling the axial current,  
and the different  weak interactions rates involving 
the T=1 1$^+$ triplet in mass 12
are consistent with one another.   
\end{abstract}
\maketitle
Quenching of spin matrix elements relative to model predictions
is a common phenomenon in nuclear structure calculations\cite{towner}.
There are a number of reasons that this occurs, including
inadequate model wave functions and the need to include meson-exchange and relativistic
corrections to the operators involved. 
For low-energy M1 $\gamma$-transitions, magnetic moments, and Gamow-Teller $\beta$-decays
effective one-body operator have been derived from calculations\cite{towner,arima}
 of the higher order corrections
to both the wave functions and the bare operators,
 and from empirical fits\cite{wildenthal} to a large body of data. 
The simplest effective one-body operators involve a scaling of the orbital and spin 
gyromagnetic ratios appearing in the magnetic operators and a scaling
of the axial vector coupling constant
for the 
weak interaction Gamow-Teller (GT) strength.

For weak interaction processes involving higher momentum transfers, such as energetic neutrino scattering 
and muon capture,
the issue of  effective operators  becomes more complicated because of the possible need for 
a momentum dependent effective coupling constants.
Siiskonen et al.\cite{sii} have examined the quenching of the axial vector coupling
by calculating corrections to the bare operator
within the shell model. They found that the quenching
factor remains approximately constant up to about 60 MeV/c, above which it becomes momentum dependent. 
Cowell and Pandharipande\cite{cowell} have calculated the momentum dependent 
quenching of the weak charge-current in nucleon matter
and found that most of the quenching is due to spin-isospin correlations induced by
one pion exchange interactions. 

A momentum independent scaling to estimate neutrino cross sections has been
used by Engel {\it et al.}\cite{engel} and Auerbach and Brown\cite{brown}.
Engel {\it et al.}\cite{engel} examined the transition to the 1$^+$ T=1 
isospin triplet in mass 12 and determined
the axial quenching factor from $\beta$-decay.
Different models that had been fitted to various combinations of the
electron scattering transverse magnetic form factor, $\beta$-decay, and $\mu$-capture
were found to give
very similar predictions for $(\nu_e,e^-)$ and $(\nu_\mu,\mu^-)$ scattering
up to momentum transfers of about 100-200 MeV/c, provided the axial vector coupling constant was
scaled separately for each model.
All of the scaled models examined were in reasonable agreement 
with the measured LSND cross sections.
More recently, Auerbach and Brown\cite{brown} have examined 
quenching of the weak axial isovector strength
in $p$-shell  nuclei within the shell model
using a single momentum-independent
quenching factor. They found that a single quenching factor worked reasonably well
for the lower momentum
transfer processes but not for
the LSND $(\nu_\mu,\mu^-)$ neutrino cross section on $^{12}$C from pion
decay-in-flight neutrinos (DIF)\cite{DIF},
where the average momentum transfer is $\sim 200$ MeV/c.
They  concluded that the failure of the model
to reproduce  
the DIF cross section with the same quenching factor needed for the 
low-momentum transfer weak processes
was evidence for a 
systematic problem with 
the LSND DIF data. 

In this brief report I discuss the  
difficulties that arise from the use of a momentum independent  
scaling factor or
effective axial coupling constant. In general, such scaling cannot be used
to describe weak 
interaction processes
involving quite different momentum transfers. 
Unless the model used is known to provide a reasonable description of the 
the semileptonic (or electromagnetic) form factor at the momentum transfers
of interest (as was the case in \cite{engel}),
estimates of high energy neutrino cross sections
from experimentally determined  cross sections
for lower energy neutrino spectra (or $\mu$-capture)
are unlikely to be reliable. 
As discussed below, such arguments provide no 
evidence for a problem with the LSND DIF  
cross sections.  Rather they suggest a short coming
in the model used.

The problem with simple scaling arguments
can be demonstrated
by examining model predictions for the transition
to the T=1 $1^+$ isospin triplet in mass 12, for which there are extensive experimental data. 
Within a $p$-shell model the one-body transition is completely described
by specifying four one-body density matrix elements (OBDMEs) and the oscillator parameter $b$.
Several authors\cite{engel, haxton,dubach,brady} have 
obtained fits to the OBDMEs and the oscillator parameter from
different combinations of the available experimental data. Comparisons
of the predictions of these fitted models
 for neutrino scattering have been presented in \cite{engel}. 

There are four basic single particle operators the contribute to the T=1 1$^+$ transition in mass 12.
In general, the $J = 1^+$ electron scattering form factor can be expressed as
\begin{equation}
F(q) = <J_f||T^{mag}(q)||J_i> =  \frac{2\sqrt{2}q}{Z M_n b}(A+By+Cy^2 + ...)e^{-y} f_{sn}(q) f_{cm}(q),
\end{equation}
where $y=(bq/2)^2$, $f_{sn}(q)$ is the single nucleon form factor,
and $f_{cm}(q)$ is the center-of-mass correction. 
The coefficients A,B,C,D,... are determined by both the structure
of the  operator $T^{mag}$ and the OBDMEs 
describing the transition. 
A similar expression holds for the axial form factor contributing to neutrino
scattering. 

If the coefficient $A$ can be determined from $\beta$-decay or some other
low-q observable (as is the case for the $1^+$ transition in mass 12),
then those weak interaction processes for which $A>>By$ are likely to be
well described by simply scaling the model form factors to $A$.
However, for processes involving momentum transfers approaching 
 $q \sim \frac{2\sqrt{-A}}{b\sqrt{B}}$  $q$-independent scaling arguments are invalid.
To correct for an inadequate model description of the higher terms in $y, \;y^2,\; ..$ in the form factors
one has to calculate explicitly a
momentum dependent effective scaling.  Alternatively, the 
model can be adjusted (or fitted) provide a good description of the (e,e') form factor up 
to momentum transfers of interest, see [6, 9-11].  
In the case of the $1^+$ transition in mass 12, the magnitude of the second term in the (e,e')
 form factor becomes equal to the first
at about 250 MeV/c. Thus, the term $By$ is an important contribution to the LSND DIF cross section, and a simple scaling of the model to reproduce the term $A$ is not sufficient. 

A simple demonstration of the problem is obtained by
considering what happens if one simply changes the oscillator parameter for a given model calculation.
This clearly has the effect of changing the relative magnitudes of the terms $A$, $By$, $Cy^2$, and therefore
of changing the predicted ratio of the different neutrino cross sections. 
For the CK(8-16) OBDMEs A=-.468 and B=.226. 
Increasing the
oscillator parameter shifts the position of the first minimum 
in the form factor and generally
shift the form factor to lower $q$. 
An oscillator parameter of $b=1.64 $ fm is needed
to fit the $^{12}$C ground state rms radius, while a larger value  
$b=1.82- 1.888 $ fm \cite{brady,dubach} is needed for the 
transverse magnetic (e,e') form factor. 
The difference in the predicted shape of the (e,e') form factor
for these two oscillator parameters for the CK wave functions is shown in Fig.1.
Also shown are the average momentum transfers involved in muon capture, the $(\nu_e,e-)$ cross section
from neutrino produced from the decay of the pion at rest (DAR), and the LSND DIF cross sections.
The large difference between the predicted and measured shape for the (e,e') form factor
makes it impossible to find a momentum independent scaling correction to the CK
prediction, especially for $b $= 1.64 fm.
It is also clear from Fig. 1 that of the 3 weak processes considered the LSND DIF cross section
is the most difficult to reproduce by simply quenching the axial vector coupling.

Table 1 lists the predicted weak interaction rates for
the CK wave functions using the two different oscillator parameters.  
For b=1.64 fm, the muon capture\cite{muon} and $(\nu_e,e^-)$ neutrino cross sections\cite{DAR} are in reasonable
agreement with experiment, being about 10$\%$ too high. However, the $(\nu_\mu,\mu^-)$ cross section
is $50\%$ too high. This is consistent with the calculations of Auerbach and Brown who use the
the same size model space, b=1.64 fm, but a different $p$-shell interaction. These latter calculations
predict the muon capture rate and the $(\nu_e,e^-)$ cross section about 60$\%$ too high and the
$(\nu_\mu,\mu^-)$ cross section about 212\% too high.
Although, the CK wave functions are considerable closer
to experiment, the two calculations show about the same level of discrepancy between the low $q$ and high $q$ 
weak processes. In strong contrast,  the CK model predictions for b=1.888 fm
provides a reasonable description
of each of the three weak interaction processes 
and the predictions agree with experiment within the quoted experimental uncertainties.
This is not surprising since the
model fits the (e,e') form factor reasonable well up 
to the average momentum transfer
involved in all three weak processes.
It does, however, simply demonstrate the increased importance of the higher order terms in eq. (1) for
the DIF cross section and hence the danger in approximating the q-dependent core-polarization corrections
to weak interaction form factors by a $q$-independent scaling.

A single effective coupling constant or scaling factor works only 
for processes involving momentum transfers where the shape of the form factors
involved is correctly predicted. 
Otherwise, the use of such a scaling factor can produce quite unreliable results, 
particularly for higher momentum transfer processes
where the shell model calculations restricted one-body weak currents
have traditionally had difficulty.
In contrast to the conclusions of Auerbach and Brown, we find no evidence for a problem
with the LSND DIF exclusive cross section. This is in agreement with the findings 
of \cite{engel,hayes}. The shortcomings of a momentum independent quenching
of the axial vector coupling constant is likely to be even worse for the LSND inclusive
DIF cross section.
There the final states are unbound and cannot be described with harmonic oscillator wave functions,
and the cross section is difficult to predict accurately \cite{hayes2}.

\bigskip
\bigskip
\vspace*{15cm}
\begin{table}

\caption{ Predicted weak interaction rates for the  $^{12}$C$\rightarrow T=1\;\; 1^+$ transitions. 
 The units are $10^{-42} cm^2$ for the $(\nu_e,e^-)$ DAR  
cross section, $10^{-40} cm^2$ for $(\nu_\mu,\mu^-)$ DIF cross section and $10^3sec^{-1}$ for muon capture.}

\begin{tabular}{|ccccc|}

\hline
& & & &\\
Interaction   & CK b=1.64 fm & CK b=1.888 fm &  Auerbach + Brown&Experiment\\
& & & &\\\hline

 $(\nu_e,e^-)$ & 9.93 &    9.5&       14.6 &      8.9$\pm$0.3$\pm$0.9 \cite{DAR} \\
 $(\nu_\mu,\mu^-)$&0.922 &  0.66&     1.4  &    0.56$\pm$0.08$\pm$0.01 \cite{DIF}  \\
$\mu$-capture& 6.41 & 5.6&    9.4 & 6.0$\pm$0.4 \cite{muon}\\  

\hline
\end{tabular}
\end{table}

 \begin{figure}
\vspace*{2cm}
\includegraphics{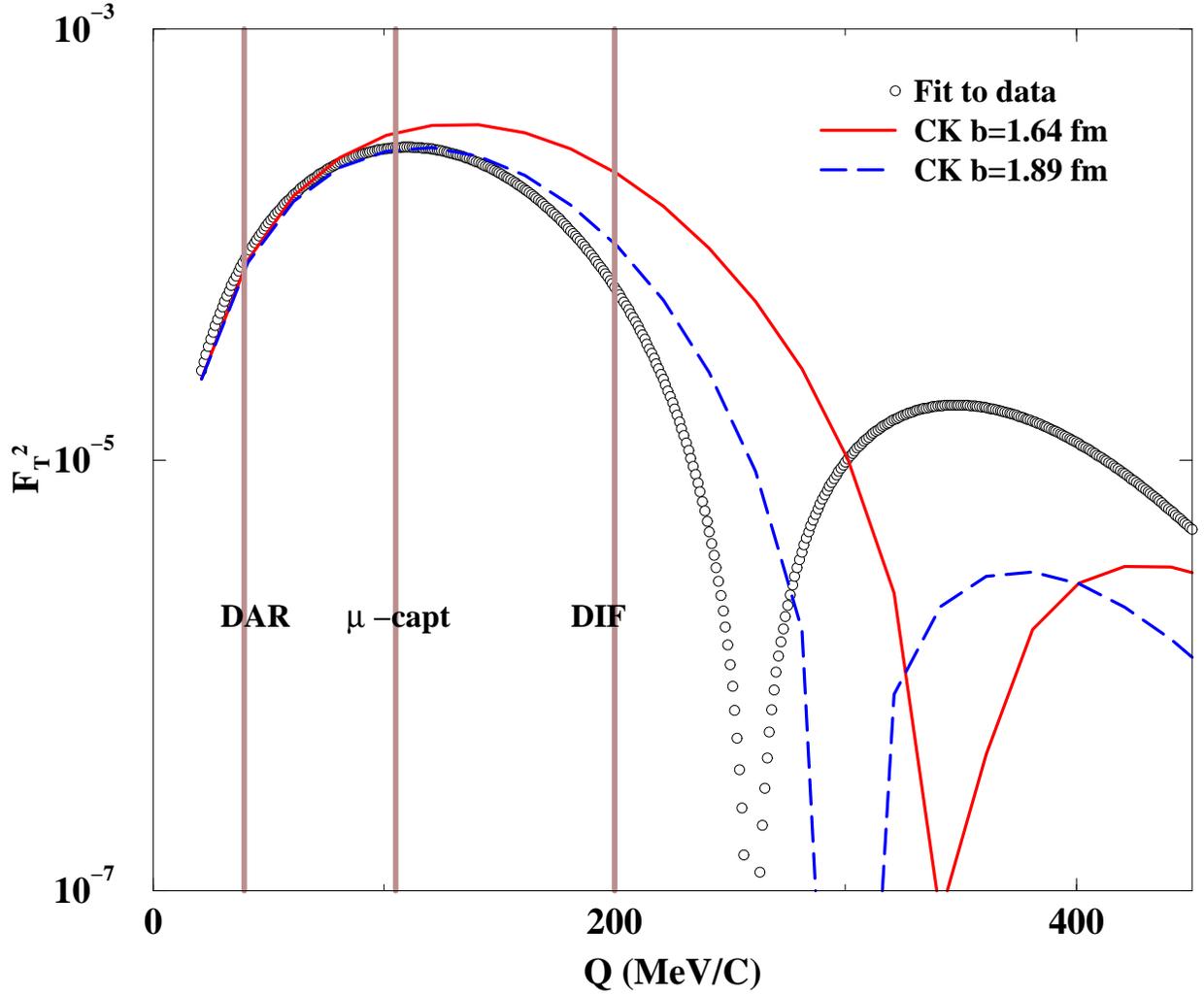} 
\caption{The transverse magnetic electron scattering form factor
for the 15.11 MeV T=1 $1^+$ state in $^{12}$C. The solid (dashed)
 curve is the Cohen-Kurath prediction using b=1.64 fm (1.89 fm)   
The vertical lines show the average momentum transfer for DAR, $\mu$-capture and DIF.  
}
\end{figure}

\end{document}